\documentclass[preprint,aps,prl,floatfix]{revtex4}
\usepackage{graphicx}
\usepackage{dcolumn}
\usepackage{bm}
\newcounter{state1}
\newcounter{state2}
\begin{document}
\title{Magnetic-Field-Induced Crossover to a Nonuniversal Regime in a Kondo Dot}
\author{Tai-Min Liu} \affiliation{Department of Physics, University of Cincinnati}
\author{Bryan Hemingway} \affiliation{Department of Physics, University of Cincinnati}
\author{Steven Herbert}
\affiliation{Physics Department, Xavier University}
\author{Michael Melloch}
\affiliation{School of Electrical and Computer Engineering, Purdue University}
\author{Andrei Kogan}
\email{andrei.kogan@uc.edu}
\affiliation{Department of Physics, University of Cincinnati}
\date{\today}
\begin{abstract}
We have measured the magnetic splitting,  $\Delta_K$, of a Kondo peak in the differential conductance of a Single-Electron Transistor while tuning the Kondo temperature, $T_K$, along two different paths in the parameter space: varying the dot-lead coupling  at a constant dot energy,  and vice versa. At a high magnetic field, $B$, the changes of  $\Delta_K$ with $T_K$ along the two paths have opposite signs, indicating that $\Delta_K$ is not a universal function of $T_K$. At low $B$, we observe a decrease in $\Delta_K$ with $T_K$ along both paths, in agreement with theoretical predictions. Furthermore, we find $\Delta_K/\Delta<1$ at low $B$ and $\Delta_K/\Delta>1$ at high $B$, where $\Delta$ is the Zeeman energy of the bare spin, in the same system. 
\end{abstract}

\pacs{72.15.Qm, 73.23.Hk, 75.20.Hr}
\keywords{Kondo, many-body phenomena, transport, non-equilibrium, single-electron transistor}
\maketitle

Universality and scaling \cite{Patashinskii:79} describe many phenomena, both equilibrium (e.g. thermodynamics of near critical fluids or ferromagnets) and non-equilibrium (transport, turbulent flow).  A famous  quantum system exhibiting universal dependence of properties on a single intrinsic energy scale (called the Kondo temperature, $T_K$)  is the Kondo singlet, a correlated ground state of a confined spin interacting coherently with delocalized electrons \cite{Hewson:93}. While the equilibrium scaling in Kondo systems is now well understood, a lot less is known about the nonequilibrium regime, i.e. when the Fermi sea near the spin confinement site is perturbed. The nonequilibrium Kondo phenomena, governed by the delicate  interplay of correlations, spin coherence, and dissipation, can be observed in Single Electron Transistors (SETs) \cite{Grobis:07,goldhaber98:NaturePaper,cronenwett98:ScienceKondo,Glazman:88,Ng:88}: electric transport in SETs is strongly affected by the Kondo effect, and the deviation from equilibrium near the spin confinement site (called quantum dot) can be precisely controlled by an externally supplied bias voltage, $V_{ds}$. One of the open questions is whether the nonequilibrium Kondo physics is universal, and, if so, whether $T_K$, as defined for an equilibrium system, remains the relevant energy scale. Very recently, Grobis et. al. \cite{Grobis:08} have shown that, at low biases and temperatures, the SET conductance $G=\partial I /\partial V_{ds}$ is a universal function of $eV_{ds}/k_B T_K$, where $e$ is the electronic charge. The splitting of the zero-bias Kondo peak in $G(V_{ds})$ in an external magnetic field $B$ \cite{meir93:nonequil} is another nonequilibrium effect predicted to exhibit universal dependence on $T_K$, with  $B/T_K$ being the scaling variable \cite{moore00:ksplit, Logan:01}.  These predictions have not been systematically tested in reported studies \cite{goldhaber98:NaturePaper,cronenwett98:ScienceKondo,Liang:02,Quay:07,Kogan:04,Amasha:05,Jespersen:06}.

In this Letter, we investigate the dependence of the Kondo peak splitting on $T_K$ at different $B$ by employing two independent parameters to tune $T_K$, rather than a single gate voltage \cite{Amasha:05}. This new approach tests whether the changes in the splitting with $T_K$ have a universal character at each $B$ independent of the observations at other values of $B$. An attractive feature of this method is that a weak dependence of $T_K$ on $B$ due to orbital effects, usually present in experiments with SETs \cite{epaps}, does not complicate data interpretation.  At low $B$, our data are qualitatively consistent with the predicted universal dependence of the splitting on   $T_K$ \cite{moore00:ksplit, Logan:01} as well as with the earlier experiment \cite{Amasha:05} but are incompatible with the universality at high $B$. The tunable parameters in this work are the energy $\epsilon_0$ of the local orbital measured from the Fermi energy of the leads, and the effective dot-lead coupling $\Gamma = \Gamma_S+\Gamma_D$, where  $\Gamma_S$ and $\Gamma_D$ are the tunneling rates from the source and drain leads onto the dot. Increasing $\Gamma$ or $|\epsilon_0 + U/2|$, where $U$ is the quantum dot charging energy,  makes $T_K$ larger \cite{Haldane:78}. When $B$ is applied,  the peak in $G(V_{ds})$ splits into two peaks with the maxima at energies $\pm\Delta_K$ relative to the Fermi energy of the leads. Theory \cite{moore00:ksplit, Logan:01} predicts that $\Delta_K/\Delta$ increases with $\Delta/k_B T_K$, as a universal function, and  therefore $\Delta_K$ decreases with $T_K$.  Here $\Delta= g \mu_B B$ is the Zeeman  splitting of a spin-degenerate orbital level in the absence of Kondo correlations, $g$ is the g-factor and $\mu_B$ is the Bohr magneton. We find a marked change in the $\Delta_K(T_K)$ dependence as $B$ increases. At low $B$, $\Delta_K$ decreases with $T_K$, as  predicted\cite{moore00:ksplit, Logan:01}. At high $B$, however, $\Delta_K$ increases with $T_K$ when we increase $\Gamma$ while keeping $\epsilon_0$ constant, but  decreases with $T_K$ when we vary $\epsilon_0$ at constant $\Gamma$. We conclude that the predicted universality with respect to $T_K$ breaks down at high $B$.

\begin{figure}
\includegraphics[width=3.8in, keepaspectratio=true]{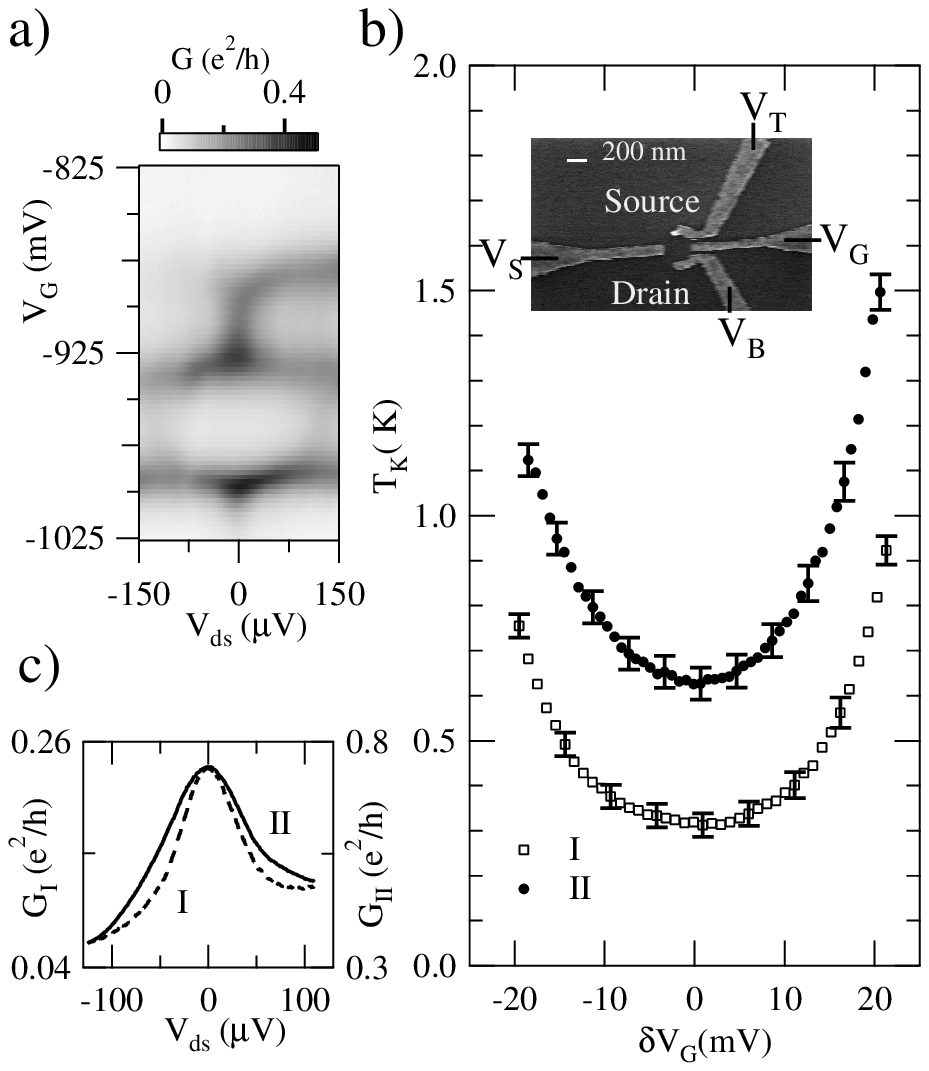}
\caption{\label{fig:G} a) Differential conductance $G$ as a function of $V_G$ and drain-source bias $V_{ds}$.  b)  T$_K$ as a function of $\delta$V$_G$ for two SET configurations.  Inset: An electron micrograph of our SET devices' gate pattern with the gate voltage labeling convention shown.  c) The Kondo peak in Configuration \Roman{state1} ($V_B=-825$ mV, $V_T=-942$ mV, $V_S=-908$ mV, and $V_G=-914$ mV) and in Configuration \Roman{state2} ($V_B=-825$ mV, $V_T=-942$ mV, $V_S=-880$ mV, and $V_G=-984$ mV). The dot occupancy is the same in \Roman{state1} and in \Roman{state2}. }
\end{figure}

\begin{figure}
\includegraphics[width=3in, keepaspectratio=true]{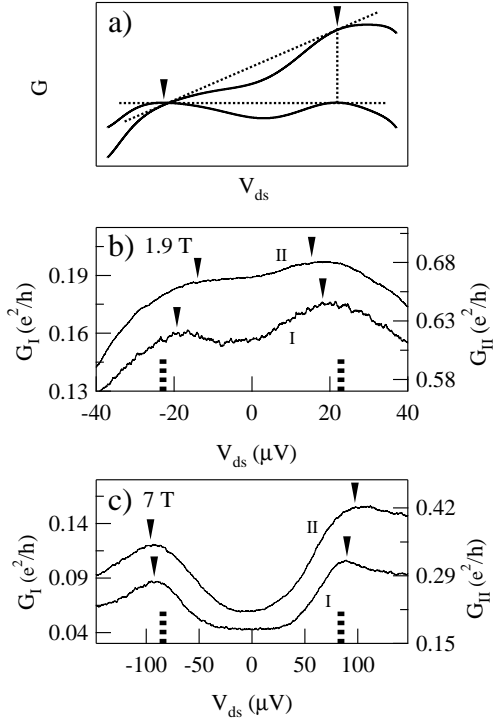}
\caption{\label{fig:data}a) Obtaining $\Delta_K$ from measured $G(V_{ds})$.  b), c) Representative conductance data in configurations \Roman{state1} and \Roman{state2} at $B$= 1.9 T (b) and $B$=7 T (c). The peak positions as determined by our procedure are marked with arrows next to the traces. The Zeeman voltage scale,  $V_{ds}=\pm \Delta/e$, is shown with short dashed lines\cite{epaps}.}
\end{figure}

To measure the differential conductance $G$, we use standard lock-in techniques with a 17 Hz, 1.9 $\mu$V RMS excitation voltage added to $V_{ds}$. Depending on the visibility of the peak splitting, we vary the acquisition time of  a single  $V_{ds}$ scan between 15 minutes and 4 hours with time constants ranging from 1 s to  30 s and, for the smallest splittings that we record, average several measurements. Reproducible splittings were observed  several months apart in measurements separated by multiple gate voltage cycles. Our SETs (Fig. \ref{fig:G}b, inset) were fabricated on a modulation-doped GaAs/AlGaAs heterostructure containing a two-dimensional electron gas (2DEG) 85 nm below the surface with the sheet density $n_{2D} =  4.8\times 10^{11}$ cm$^{-2}$ and mobility $\mu \ge 5\times 10^{5}$ cm$^2$/V\,sec. The metal  gates (5 nm Ti/ 20 nm Au) were patterned via e-beam lithography followed by lift-off.  The tunneling rates $\Gamma_S$ and $\Gamma_D$ are controlled by the pairs ($V_S$, $V_T$) and  ($V_S$, $V_B$) of gate voltages. We note that  $V_S$ controls both $\Gamma_S$ and $\Gamma_D$. A deviation, $\delta \epsilon$, of the  dot energy from the middle of the Coulomb valley is a linear combination of the changes in the gate voltages and $V_{ds}$:
\begin{eqnarray*}
(\delta \epsilon)/e = \alpha_S (\delta V_S) + \alpha_T (\delta V_T) + \alpha_B (\delta V_B)\\
+ \alpha_G (\delta V_G)+ \alpha_{ds} (\delta V_{ds})
\end{eqnarray*}
where $\alpha_i = C_i/C_{total}$ are the mutual capacitances between the dot and the device gates and leads, expressed as  fractions of the total capacitance of the dot. Using standard techniques, we find  $U=1.4\pm0.05$ meV, $\alpha_{ds} = 0.4\pm0.03$, $\alpha_G = 0.024\pm0.0018$, $\alpha_S/\alpha_G = 2.4$, $\alpha_T/\alpha_G = 2.5$, and $\alpha_B/\alpha_G = 2.0$. We estimate the actual dot diameter to be $\sim 0.13\, \mu$m, which gives an average orbital level spacing of $\sim 540\,\mu\mbox{eV}$. The external magnetic field $B$ is aligned parallel to the 2DEG plane to $\pm 1^\circ$. Using the spin-flip cotunneling spectroscopy method \cite{Kogan:04}, we obtain the base electron temperature for our dilution refrigerator (Leiden Cryogenics)  $ T_{el}\le 55$ mK and the heterostructure $g-$factor $|g|=0.2073\pm0.0013$, which gives $\Delta = 12.02$  $\mu$eV/T. Fig. \ref{fig:G}a shows a Kondo valley between $V_G =-930$ mV and $V_G=-870$ mV flanked by Kondo-free, even-occupied valleys.  The Kondo temperature at different gate voltages,  obtained from measurements of the temperature dependence of the Kondo conductance \cite{epaps},  is plotted in Fig. \ref{fig:G}b. The zero-bias mid-valley Kondo peak (Fig. \ref{fig:G}c) is shown for two device configurations: \Roman{state1} with $\Gamma$=0.53 meV, $T_{K}$ = 0.3 K, and \Roman{state2} with $\Gamma$ = 0.7 meV and $T_{K}$ = 0.63 K. We note that, despite the difference in $V_G$, both \Roman{state1} and \Roman{state2} correspond to the middle of the Kondo valley, i.e. $\epsilon_0 \sim -U/2$ for both. The choice of the  \Roman{state1} and \Roman{state2} configurations is a compromise between minimizing thermal effects, which favors higher $\Gamma$ and $T_K$, and avoiding mixed-valence corrections \cite{goldhaber98:PRL}, which limits the largest usable $\Gamma$  to $|\epsilon_0|/\Gamma \ge 0.5$ \cite{goldhaber98:PRL}. When sweeping between \Roman{state1} and \Roman{state2}, the relative electron temperature $T_{el}/T_{K}$ varies  from 0.18 to 0.09, and  $|\epsilon_0|/\Gamma>1$ is maintained.  To obtain $T_K$, we fit the temperature dependence $G(T)$ in each configuration to the empirical form $G(T)= G_0[1+(2^{1/0.22}-1)(T/T_K)^2]^{-0.22}$ \cite{goldhaber98:PRL}. We estimate $\Gamma$ independently from the width of the charging peak with the Kondo effect thermally suppressed, and from fits of  $T_K(V_G)$ to the Haldane function \cite{Haldane:78}, and find good agreement. Using the height of the Kondo peak to estimate the device asymmetry,  $\Gamma_S/\Gamma_D$, we get 24 (\Roman{state1}) and 7  (\Roman{state2}).  To obtain $\Delta_K$ from measured $G(V_{ds})$ (Fig. \ref{fig:data}a)\cite{epaps}, we first fit the data near each maximum to an analytical function $G^{fit}(V_{ds})$. To account for the slight peak height difference, we subtract a linear background from the fits to equalize the  maxima and then obtain $V_{ds}$ for the left and the right peaks by solving $d\,G^{fit}/d\,V_{ds}=0$. $\Delta_K/e$ is taken as half the difference between these  $V_{ds}$ values. We find that the result varies by no more than $\sim$ 2 $\mu$V for different sensible choices of the background slope.

\begin{figure}
\includegraphics[width=3in, keepaspectratio=true]{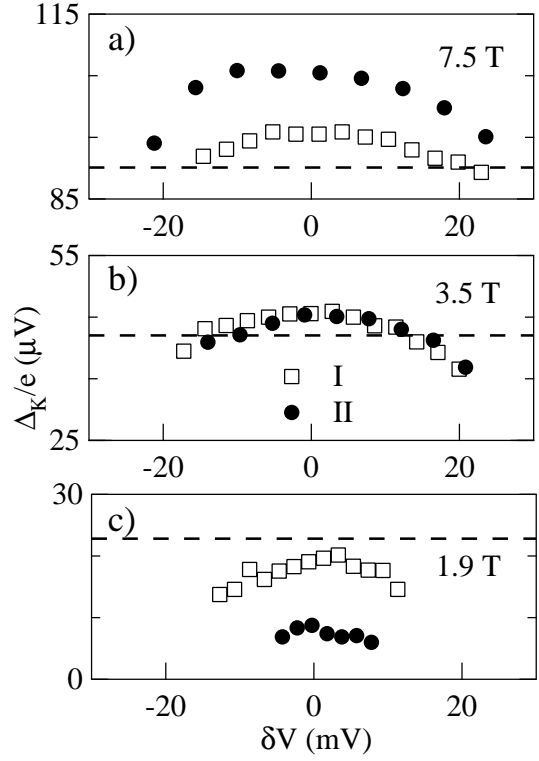}
\caption{\label{fig:vsg} Variation of $\Delta_K$  with gate voltage $V_G$ at different values of $B$ field.  $\delta V_G=0$ corresponds to configuration  \Roman{state1} (open squares) and  \Roman{state2} (filled circles) as defined in the caption of Fig. 1. The Zeeman bias voltage scale, $\Delta/e$, for each $B$ is marked with a dashed line. $T_K$ in \Roman{state1} and \Roman{state2} is 0.3 and 0.63 K, respectively}
\end{figure}

\begin{figure}
\includegraphics[width=3in, keepaspectratio=true]{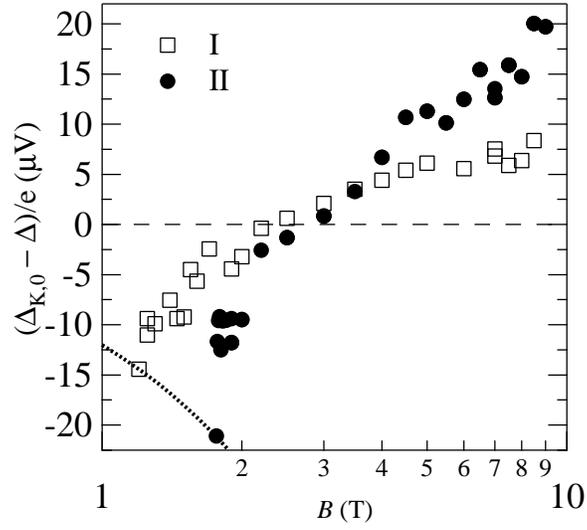}
\caption{\label{fig:vsb}Deviation of mid-valley splitting $\Delta_{K,0}$ from Zeeman energy $\Delta$  as a function of magnetic field.  The horisontal dashed line corresponds to $\Delta_{K,0}=\Delta$. The dotted line corresponds to zero peak splitting: $\Delta_{K,0}$=0\cite{epaps}.}
\end{figure}

\begin{figure}[b]
\includegraphics[width=3in, keepaspectratio=true]{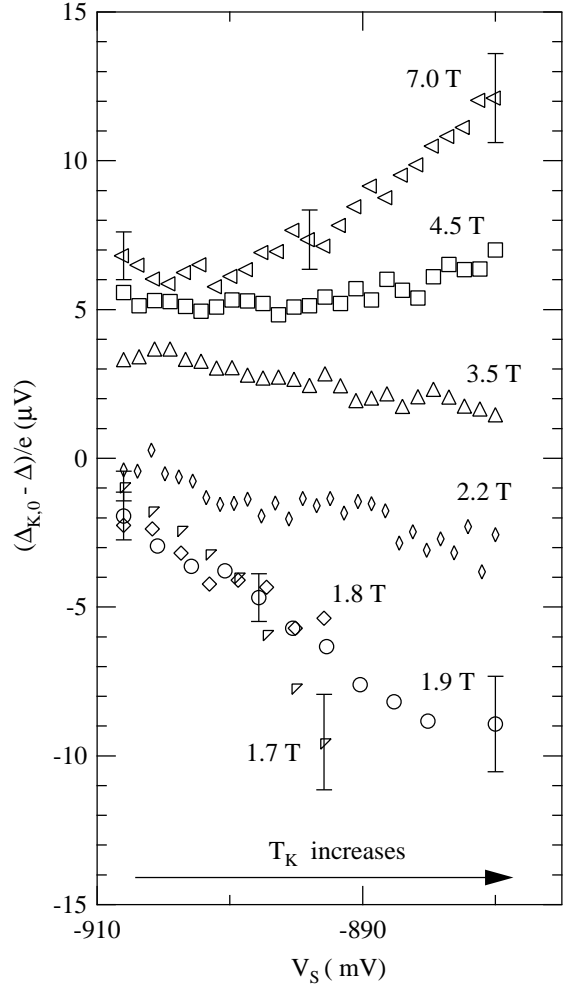}
\caption{\label{fig:dev}Deviation $\Delta_{K,0}-\Delta$ as a function of $V_S$ along a constant $\epsilon_0$ path. Both $V_S$ and $V_G$ are swept.  The dot energy is set in the middle of the Kondo valley and $\alpha_S(\delta V_S) + \alpha_G(\delta V_G)=0$ is maintained while $V_S$ is varied.}
\end{figure}

We open the discussion of the results by examining how $\Delta_K$ changes with $\epsilon_0$ at constant $\Gamma$. Starting with all gate voltages set to  \Roman{state1} (see Fig. 1, caption), we scan $V_G$, and then repeat the experiment with \Roman{state2} as the starting point.  Fig. \ref{fig:vsg} presents plots of $\Delta_K$ as functions of the deviation, $\delta V_G$, of $V_G$ from the value that corresponds to  \Roman{state1} (open squares) and \Roman{state2} (filled circles). At all  $B$, $\Delta_K$ decreases as $V_G$ is tuned away from the center of the valley, which corresponds to an increase in $T_K$. This agrees with the earlier observations \cite{Amasha:05}. Next, we fix the dot energy in the middle of the valley  and focus on the changes of the splitting with $\Gamma$.
A detailed dependence of the mid-valley splitting $\Delta_{K,0}$ on $B$ for the configurations \Roman{state1} and \Roman{state2} is presented in Fig. \ref{fig:vsb}.  First, we note that the lowest magnetic field  at which the Kondo peak shows detectable splitting increases with  $T_K$. Introducing the corresponding Zeeman scale $\Delta^{onset}$, we find  $\Delta^{onset}= 0.55\, k_B T_K$  (\Roman{state1}) and $\Delta^{onset}= 0.4 \,k_B T_K$ (\Roman{state2}). These are in reasonable agreement  with the prediction $\Delta^{onset}=0.5 \,k_B T_K$ \cite{costi00:kondoh} and  are somewhat lower than the previously reported $\sim 0.86 \,k_BT_K$ \cite{Amasha:05}, and $\sim 0.8 \,k_BT_K$ \cite{Quay:07} possibly due to a heavy signal averaging and a lower  relative electron temperature ($\sim$1/15 to 1/6 in the present work vs ~1/6 in \cite{Quay:07},  and ~ 1/3 in \cite{Amasha:05}). Near the onset, the data show a pronounced suppression  $\Delta_{K,0}<\Delta$, consistent with, but much stronger, than in the earlier report by Quay et. al. \cite{Quay:07} who used a  carbon nanotube-based SET. We note that in the earlier experiments with  heterostructure-based SETs \cite{Kogan:04,Amasha:05}, $\Delta_{K,0}<\Delta$ was not observed. As $\Delta$ increases above $\sim k_B T_K$,  $\Delta_{K,0}<\Delta$ is replaced with the $\Delta_{K,0}>\Delta$ regime. To our knowledge, such a transition at a finite $B$ has not yet been reported, although $\Delta_K > \Delta$  was  previously observed experimentally  \cite{Kogan:04,Amasha:05,Jespersen:06} \footnote{The authors of ref. \cite{Jespersen:06}  point out that their data are consistent with $\Delta_K/\Delta=1$  if one defines $\Delta_K$ as the position of the steepest point, rather than the maximum of, $G(V_{ds})$, as suggested by Paaske et. al. \cite{Paaske:04}}  ,found theoretically in the very recent calculations by Hong and Seo \cite{Hong:08}, and also, for the $B>>k_BT_K$ regime, predicted  by the perturbative method described by Paaske et. al. \cite{Paaske:04}.  At yet higher $B$, our $\Delta_{K,0}$  data for the more open, higher $T_K$ configuration  \Roman{state2} exceed those for the lower $T_K$ configuration \Roman{state1}. This is opposite of what we find in the fixed $\Gamma$ experiments (Fig. 3 and its discussion), in which $\Delta_{K,0}$ decreases with $T_K$ regardless of the magnitude of $B$.   To examine the $\Delta_K(T_K)$ dependence at fixed energy in more detail, we follow a constant $\epsilon_0$, variable $\Gamma$ path starting in the middle of the Kondo valley (Fig. \ref{fig:dev}). In the beginning of each sweep, the device is set to \Roman{state1}. Then, both $V_S$ and $V_G$ are swept simultaneously so as to keep $\alpha_S(\delta V_S)+\alpha_G(\delta V_G)=0$, and thus maintain $\epsilon_0 \sim -U/2$. The changes in $T_K$ during such a sweep come from the changes in $\Gamma$ only, and the device is being tuned continuously from \Roman{state1} to \Roman{state2}. To verify that the presence of the magnetic field does not reverse the expected increase of  $\Gamma$ with $V_S$, as may occur, for example, due to an accidental scattering by impurities near the tunnel barriers, we have compared the charging peak widths in configurations \Roman{state1} and \Roman{state2} measured at zero bias with a 9 Tesla  magnetic field applied. We found the ratio of the widths to be 0.6. This is comparable to the ratio of $\Gamma$ values in \Roman{state1} and \Roman{state2} at zero magnetic field (0.78), and indicates that the configuration \Roman{state2} remains stronger coupled to the leads than configuration \Roman{state1} even at the highest available magnetic field. At low $B$, we observe  $\Delta_{K,0} < \Delta$ and $\Delta_{K,0}$ decreasing with increasing $V_S$ (and also $\Gamma$ and $T_K$). At fields larger then $\sim 4$ T, the opposite occurs:  $\Delta_{K,0}>\Delta$ and increases as  $\Gamma$ and $T_K$ increase. Thus, at high $B$, scaling with $T_K$ breaks down: changes of $\Delta_K$ with $T_K$ in the constant energy and in the constant $\Gamma$ experiments have opposite signs. Interestingly, both the high $B$ and the low $B$ trends shown in Fig. 5 agree qualitatively with the $\Delta_K \rightarrow \Delta$ behavior expected for the limit of small $\Gamma$ and $T_K$ \cite{moore00:ksplit,Costi:03,Logan:01}. 

In summary, we have measured $\Delta_K$ as a function of $B$ and two parameters, $\epsilon_0$ and $\Gamma$, that influence $T_K$. At a sufficiently large $B$, a crossover occurs to a regime in which a universal dependence of $\Delta_K$ on $T_K$ is qualitatively inconsistent with the data. In addition, we  observe both $\Delta_K<\Delta$ (low $B$) and $\Delta_K>\Delta$ (high $B$) regimes in a single SET system, and find that the transition between the two regimes occurs at $B$ values comparable to those for the crossover.  

The research is supported by the NSF DMR award No. 0804199 and by University of Cincinnati. We are grateful to M. Jarrell, R. Serota and M. Ma for helpful discussions. We thank K. Herrmann, A. Maharjan  and M. Torabi for their help with the transport circuit construction and sample fabrication, and J. Markus and R. Schrott for technical assistance. S. H. acknowledges the support from the John Hauck Foundation. T.-M. L. acknowledges SET fabrication support from the Institute for Nanoscale Science and Technology at University of Cincinnati.
\bibliographystyle{apsrev}

\newpage
\end{document}